\documentstyle[aps,twocolumn,pra,epsf,eqsecnum]{revtex} 

\begin{document} 

\title{Testing the Symmetrization Postulate on Molecules with Three
  Identical Nuclei}

\author{Giovanni Modugno$^1$ and Michele Modugno$^2$}
\address{$^1$ INFM and LENS, 
Largo E. Fermi 2, I-50125 Firenze, Italy.}
\address{$^2$ INFM and Dipartimento di Fisica, Universit\`a di Firenze, \\
Largo E. Fermi 2, I-50125 Firenze, Italy.}

\date{\today}
\draft
\maketitle

\begin{abstract}
We propose an experiment to look for possible small violations of the
symmetrization postulate of Quantum Mechanics, in systems composed by
three identical particles. Such violations could be detected by
investigating the population of particular roto-vibrational states of
symmetrical molecules containing three identical nuclei. We discuss
the symmetry properties of such states, and the implications of the
symmetrization postulate and of the spin-statistics. A high
sensitivity spectroscopic investigation on simple molecules such as
SO$_3$, BH$_3$ and NH$_3$ could lead to {\it the first test of the
symmetrization postulate} for spin-0 and spin-1/2 nuclei. 
\end{abstract} 
\pacs{03.65.Bz, 33.15.Bh, 87.64.Je} 

\section{Introduction}

In the latest years, the possibility of small violations of the
symmetrization postulate (SP) and the spin-statistics connection has been
addressed from both the experimental and the theoretical point of
view. The SP is at the basis of the
quantum-mechanical description of systems composed by identical
particles, asserting that only wave functions completely symmetric or
antisymmetric in the permutation of the particles labels can describe
physical states. In principle there would be no arguments against the
existence of states with different symmetries, although they lack some
of the properties which are peculiar of the completely symmetric and
antisymmetric states. The only strict requirement that can be derived
in a formal way in Quantum Mechanics is the so called {\it superselection
rule}, which forbids transitions between different symmetry
classes. \par

In addition, the experimental observation seems to indicate so far a
well defined connection between the spin of the particles and the
symmetry properties: half-integer spin particles can be described only
by antisymmetric wave functions, while integer-spin particles by
symmetric wave functions only. According to their behavior in a
statistical ensemble, the two kinds of particles are indicated
respectively as fermions and bosons. \par

Although both the SP and the spin-statistics connection seem to hold
in the physical world, there is great interest in investigating
possible very small violations, not detected so far. In case of
two-particles systems, only the symmetrical and antisymmetrical states
can be defined, and therefore the SP is redundant. As a matter of
fact, the experimental tests of symmetry reported so far have been
limited to the investigation of states containing only two identical
particles, looking for violations of the usual connection between
spin and statistics. In particular the latest high-sensitivity
experiments of this kind have been performed on both fermionic
systems, namely electrons, and bosonic systems, such as photons and
oxygen nuclei. In the case of electrons, the existence of totally
symmetric states in the spectrum of atomic helium has been searched
\cite{deilamian}, and the negative result of such investigation has
led to the establishment of a bound to the degree of violation of the
spin-statistics, {\em i.e.} to the presence of exchange-symmetric electrons. 
A different experiments was performed on a metallic Cu sample
\cite{ramberg}, looking for the presence of any of the electrons in
the conduction band with no symmetry requirements with respect to the
electrons in the already filled shells. This can be considered a
one-particle test of the identity of the electrons, which is even a
more general property of the particles, but not a test of their
symmetry. Results which are conceptually identical to the first case
have been obtained for bosonic particles, by searching for an
exchange-antisymmetric two-photons transition in atomic barium
\cite{demille99}, and for rotational states in molecular
oxygen \cite{deangelis,hilborn96,gianfrani} and carbon dioxide
\cite{modugno} which are antisymmetrical in the exchange of O nuclei. 
The principle of all these experiments was to look for particular
atomic or molecular transitions, whose probability would be zero
according to the usual spin-statistics, by using high sensitivity
spectroscopic detection techniques. The extrapolated bounds on the
relative weight of the wrong statistics are almost at the level of
10$^{-9}$ \cite{modugno}. Reviews of different experiments which have
been proposed, and interpretations of the results of various
measurements as tests of the statistics, can be found in
\cite{greenberg89,lamoreaux,tino99}. \par

Theoretical studies of possible violations of the usual symmetry properties
have focused on systems composed by a non-fixed number of particles, 
which can be treated within the Quantum-Field Theory formalism. In such
context the spin-statistics connection is no more a postulate dictated
by the experimental observation, but it can be rigorously proved
through the spin-statistic theorem\cite{pauli40,spinstat}, the main
assumption being the validity of the SP. Several theories allowing for
small violations of such postulate have been proposed
\cite{greenberg94}, and one of the most successful is based on the
$q$-mutator algebra \cite{greenberg90}, in which deformed bilinear
commutation relations are used in place of the ordinary ones. In this
way it is possible to define exotic symmetries as a smooth interpolation
between the symmetric and antisymmetric ones, or sta\-ti\-stics
interpolating between the Bose and Fermi cases. By assuming the
validity of this algebra it has also been possible to translate the
bound on the spin-statistics for oxygen nuclei \cite{modugno}, into a
bound on the statistics of the nucleons composing the nuclei
\cite{greenberg99} or, more generally, to establish a relationship
between constraints on different kinds of
particles\cite{greenberg98}. \par

To summarize, so far experiments have been devoted to search possible
violations of the spin-statistics in simple systems, while theories
allowing for violations of the more fundamental SP were focused on
much more complex systems, which cannot be easily investigated
experimentally. We note that no rigorous theoretical study of a simple
quantum-mechanical system allowing for violations of the SP has
been reported so far, although possible experiments on multi-particle
molecular systems have been indicated \cite{deangelis,tino99}. \par 

In this paper we propose an experiment to be performed on molecular
systems containing three identical nuclei, to search for possible
violations of the SP in a system described within the
quantum-mechanical theory. For this purpose, in Section
\ref{sect:general} we briefly recall the concept of identity of
particles, and the permutation properties of identical particles; we
also recall that at least three particles are needed to define
symmetries different from the usual ones. Then, in Section
\ref{sect:forbid}, we discuss the properties of particular rotational
states of symmetrical plane molecules with three spin-0 nuclei, which
are forbidden by the SP, and with three spin-1/2 nuclei, which are
forbidden by both the spin-statistics connection and the SP.  We show
that, nevertheless, such states have the proper symmetry to be defined
consistently in a more general quantum-mechanical theory which does
not include the SP.  In Section \ref{sect:test} we briefly discuss how
to test the possible small symmetry violations.  Finally, in Section
\ref{sect:proposal}, we propose a high-sensitivity spectroscopic
investigation to be performed on selected transitions of SO$_3$
molecule, and of BH$_3$ and NH$_3$ molecules, which best represent the
prototypal molecule respectively for the spin-0 and spin-1/2 cases
treated above. In the latter case, such states could be investigated
in search of small violation of the statistics or of the symmetry of
protons.

\section{The indistinguishability of identical particles 
and the symmetrization postulate}
\label{sect:general}

\subsection{General discussion}

The indistinguishability of identical particles is one of the principles
lying at the basis of Quantum Mechanics. It states that given a system of
particles belonging to the same species ({\em i.e.} which have the same
physical properties, like mass, charge, spin, etc.), a permutation of such
particles cannot lead to any observable effect \cite{messiah}. To discuss the
implications of such principle, here we briefly recall some properties of
a system containing $N$ identical particles (for a general discussion 
see for example Ref. \cite{messiah,caldirola,sakurai}). 

To each observable of the system is associated an hermitian operator
$\hat{A}$, which can be written as
\begin{equation}
\hat{A}=A(\{\hat{O}_1\}, \cdots \{\hat{O}_N\})\equiv\hat{A}(1, \cdots, N)
\end{equation}
where $\{\hat{O}_j\}$ is a set of single-particle observables. In the same
way, the vector
\begin{equation}
|1, 2, \cdots, N\rangle 
\end{equation}
represents a state where the first particle is characterized by a set of
quantum numbers $\{1\}$, the second by $\{2\}$, and so on. Moreover we can
define the permutation operators as
\begin{equation}
\hat{P}_{j_1, j_2, \cdots, j_N}f(1, 2, \cdots, N)=f(j_1, j_2, \cdots, j_N)
\end{equation}
and these operators form the {\em symmetrization group} $S_n$. 
Therefore, since a permutation of identical particles cannot be observed in
any experiment, all quantum observables must be permutation-invariant
\begin{equation}
[\hat{P}, \hat{A}]=0
\end{equation}
and since the evolution operator $\hat{U}(t)$ is related to the Hamiltonian
of the system $\hat{H}$, which is a physical observable, the above condition is
fulfilled at any instant of time
\begin{equation}
[\hat{P}, \hat{H}]=0\qquad \Longrightarrow \qquad [\hat{P}, \hat{U}(t)]=0. 
\end{equation}
An important consequence of these relations is that the Hilbert space ${\cal
H}$ can be written as a direct sum of hortogonal subspaces, each one invariant
under the permutation group, the hortogonality being preserved along the
evolution of the system, and following a physical measurement. 
This result can be stated as a {\em super-selection rule} which forbids
transitions between states transforming under inequivalent representations of
the permutation group. 

To investigate in detail
the symmetry properties of a generic state of the system under permutation, 
it is convenient to introduce the operator of a {\em cyclic permutation}
\begin{equation}
\hat{P}_{(k_1, k_2, \cdots, k_K)}f(k_1, k_2, \cdots, k_K)=f(k_2, \cdots, k_K, k_1)
\end{equation} and the exchange operators of two particle, $\hat{P}_{(j, k)}$
(they are a particular case of the former), which are hermitian and unitary, with
eigenvalues $\epsilon_{j, k}=\pm 1$. 

Each generic permutation can be written either as the product of exchange
between two particles, or of cyclic
permutation of distinct elements, in the form 
\begin{equation} 
\hat{P}_{cyclic}\equiv
\hat{P}_{(k_1, \cdots, k_p)
(k_{p+1}, \cdots, k_{p+q}) (k_{p+q+1}, \cdots, k_{p+q+r})\cdots
 } \end{equation} with the only condition that the sum of each cycle
length must be equal to $N$ \begin{equation}
p+q+r+\cdots = N
\end{equation}
(here we adopt the convention $p\ge q\ge r\ge \cdots$). The properties of
transformation of such operators under permutation do not change the length of
the cycles, and therefore all the cyclic permutation with the same
values of $p, q, r, \cdots$ are said to form a {\em class}. Obviously there are
as many classes as many are the possible decomposition of $N$ as sum of
integer numbers. 

From the properties of the exchange operators $\hat{P}_{(j, k)}$ it follows
that the permutation operators admit only two common eigenstates, 
with the eigenvalues $\epsilon_{j, k}$ all equal to $1$ or $-1$. In fact, the
eigenvalue equation for the exchange operators $\hat{P}_{(j, k)}$ is 
($\epsilon_{j, k}=\pm1 $)
 \begin{equation}
\hat{P}_{(j, k)}f(1, 2, \cdots, N)=\epsilon_{j, k}f(1, 2, \cdots, N)
\end{equation} 
then, from the relation
\begin{equation}
\hat{P}_{(j, k)}=\hat{P}_{(2, k)}\hat{P}_{(1, j)}\hat{P}_{(1, 2)}\hat{P}_{(2, k)}\hat{P}_{(1, j)}
\end{equation} 
it follows
\begin{equation}
\epsilon_{j, k}=\epsilon_{1, 2}\epsilon_{2, k}^2\epsilon_{1, j}^2=\epsilon_{1, 2}. 
\end{equation} 
These eigenvalues $\epsilon_{j, k}=\pm1$ correspond
respectively to vectors of state completely symmetric or antisymmetric under
the exchange of two particles, 
and we indicate the subspaces spanned by these vectors by ${\cal H}_+$ and 
${\cal H}_-$ respectively. 
For $N=2$ the Hilbert space can
be decomposed exactly as ${\cal H}={\cal H}_+\oplus{\cal H}_-$, {\em i.e.} only
symmetric or antisymmetric states are possible. For $N>2$ this is no more true,
and we have
\begin{equation}
{\cal H}={\cal H}_+\oplus{\cal H}_-\oplus{\cal H}'_1\oplus{\cal
H}'_2\oplus\cdots, \end{equation} 
where ${\cal H}'_j$ are permutation-invariant subspaces. 
Contrarily to ${\cal H}_+$
and ${\cal H}_-$, which have a definite symmetry
 and are one-dimensional representation of the 
permutation group (for a given set of
quantum numbers $\{1, 2, \dots, N\}$ there exist only one possible combination which
gives symmetric or antisymmetric states), the subspaces ${\cal H}'_j$ do not
posses a definite symmetry and have dimension greater that one. This is
the reason why one usually assumes the {\em symmetrization postulate} (SP), 
requiring physical states to be either symmetric or an\-ti\-sym\-me\-tric. 
Nevertheless, there is no stringent reason to exclude {\em a priori} the
possibility of physical states not obeying SP, since their presence do not
violate any basic principle of Quantum Mechanics. The only difference that
one has to take into account is the lack of a correspondence
between physical states and vectors (modulo a phase factor), since the subspaces
${\cal H}'_j$ have dimension greater that one and do not admit a complete set
of mutually commutating physical observables. In fact, let us imagine to
have a common eigenvector $|u\rangle$ of a complete set of operators $\{O_i$\} in
${\cal H}'_j$. Since $|u\rangle$ can not be eigenvector of all the permutation
operator, there must exist some permutation such that
$\hat{P}_{j_1, j_2, \cdots, j_N}|u\rangle$ is linearly independent of $|u\rangle$, 
and therefore ${\cal H}'_j$ must have dimension greater that one, and the set
of operators $\{O_i\}$ can not be complete ({\em i.e.} the state of the system
is not completely determined by the set of eigenvalues $\{o_i\}$). 

In the following we work out in more detail the permutation properties 
for the special case n=3 \cite{wuki}.

\subsection{A system of 3 identical particles}

The symmetrization group $S_3$ is a non-abelian group of order 6, 
formed by the $3!$ cyclic permutations of the labels $(1, 2, 3)$, 
which belong to three distinct classes
\begin{equation}
P_{(1)(2)(3)}= I
\end{equation}

\begin{equation}
P_{(2, 3)};\quad
P_{(1, 3)};\quad
P_{(1, 2)}
\end{equation}

\begin{equation}
P_{(1, 2, 3)}=P_{231};\quad
P_{(3, 2, 1)}=P_{312}. 
\end{equation}

\begin{figure}[hb]
\setlength{\unitlength}{1pt}
\begin{center}
\begin{picture}(120, 120)(-60, -40)
\put(0, 57.7){\line(3, -5){51}}
\put(0, 57.7){\line(-3, -5){51}}
\put(-50, -29){\line(1, 0){100}}
\put(0, 60){\line(0, -1){89}}
\put(8, -25){\makebox(0, 0)[r]{$a$}} 
\put(-50, -29){\line(5, 3){75}} 
\put(20, 17){\makebox(0, 0)[r]{$c$}} 
\put(50, -29){\line(-5, 3){75}}
\put(-15, 17){\makebox(0, 0)[r]{$b$}} 
\put(0, 57.7){\circle*{5}}
\put(0, 65){\makebox(0, 0)[r]{$1$}} 
\put(50, -29){\circle*{5}}
\put(60, -30){\makebox(0, 0)[r]{$2$}} 
\put(-50, -29){\circle*{5}} 
\put(-55, -30){\makebox(0, 0)[r]{$3$}} 
\end{picture}
\end{center}
\caption{The symmetry operation on this triangular configuration 
form the {\em dihedral group} $D_3$, and
are: (i) the identity transformation, (ii) reflection about the axes 
$a$, $b$, $c$ and (iii) rotation around the center by angles 
$2\pi/3$ and $-2\pi/3$.}
\label{d3}
\end{figure}
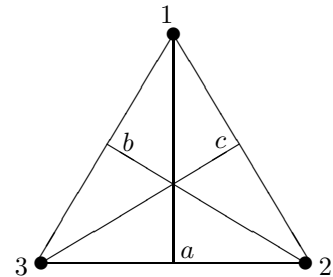

This group
is isomorphic to the {\em dihedral group} $D_3$, generated by the 
symmetry transformation shown in Fig. \ref{d3}. In fact, 
reflection about each axis ($a, b, c$) is equivalent to a two label 
permutation $P_{(i, j)}$, while rotations around the center by angles 
$2\pi/3$ and $-2\pi/3$, lead to cyclic permutation 
of all the three labels. The visualization of the group $S_3$ by its 
association to the geometrical symmetries of $D_3$ is very helpful 
since it is indeed this realization on a physical system 
(a molecule in our case) that we are looking for.

The permutation operators can be written explicitly in a matrix form, on the
basis formed by the $3!$ permutations of the vector $|1, 2, 3\rangle$. 
For example, by defining
\begin{eqnarray}
|1, 2, 3\rangle \;\equiv\;|1\rangle \\
|1, 3, 2\rangle \;\equiv\;|2\rangle \\
|2, 1, 3\rangle \;\equiv\;|3\rangle \\
|2, 3, 1\rangle \;\equiv\;|4\rangle \\
|3, 1, 2\rangle \;\equiv\;|5\rangle \\
|3, 2, 1\rangle \;\equiv\;|6\rangle 
\end{eqnarray}
we obtain
\begin{equation}
P_{(1, 2, 3)}\, =\, \pmatrix{0&0&0&0&1&0\cr0&0&1&0&0&0\cr0&0&0&0&0&1\cr
1&0&0&0&0&0\cr0&0&0&1&0&0\cr0&1&0&0&0&0}\, , 
\end{equation}
and
\begin{eqnarray}
P_{(3, 2, 1)}\, =\, \pmatrix{0&0&0&1&0&0\cr0&0&0&0&0&1\cr0&1&0&0&0&0\cr
0&0&0&0&1&0\cr1&0&0&0&0&0\cr0&0&1&0&0&0}\, . \\
\quad\nonumber
\end{eqnarray}
This matrix representation of the group $S_3$ can be reduced, since it
must contain a number of irreducible representation equal to the
number $n_c$ of distinct classes ($n_c=3$).  Moreover, since this is
the so called {\em regular representation}, each irreducible
representation must appear a number of times equal to its
dimensionality. From general results \cite{wuki} we know that these
representations are the symmetric and antisymmetric ones (which are
1-dimensional), and that there must be also a 2-dimensional
representation occurring two times.  A formal way to systematically
reduce the regular representation of $S_n$ into its irreducible
components would be by using the {\em Young diagrams} \cite{wuki}.
Here we work out the S$_3$ case with more heuristic arguments, as
follows.

First of all, we can define the operators $S$ and $A$, which are respectively
the projectors on the subspaces $\cal H_+$ and $\cal H_-$
\begin{eqnarray}
S&=&\big[
I+P_{213}+P_{312}+P_{132}+P_{312}+P_{231}\big]/6\\
A&=&\big[
I-P_{213}-P_{312}-P_{132}+P_{312}+P_{231}\big]/6\, . 
\end{eqnarray}
Each of them has therefore only one eigenvector associated to a non-zero
eigenvalue, which are respectively
\begin{eqnarray}
|s\rangle & \equiv & \Big[\, |1\rangle +|2\rangle +|3\rangle +
|4\rangle +|5\rangle +|6\rangle \, \Big]/\sqrt{6}\\
|a\rangle & \equiv&\Big[-|1\rangle +|2\rangle +|3\rangle -
|4\rangle -|5\rangle +|6\rangle \, \Big]/\sqrt{6}\, . 
\end{eqnarray}
and are the standard symmetric and antisymmetric states. 

Then, in order to generate the two 2-dimensional irreducible representations
 of $\cal H'$, we can try to diagonalize
the permutation operators $P_{(1, 2, 3)}$ and $P_{(3, 2, 1)}$. 
The eigenvectors of the former, 
 and their corresponding eigenvalues $\lambda$, are
\begin{eqnarray}
&\cases{
 |v_1\rangle\equiv\lambda_-|2\rangle\, +\, \lambda_+|3\rangle\, +\, |6\rangle & \cr
& \cr
 |v_2\rangle\equiv\lambda_+|1\rangle\, +\, \lambda_-|4\rangle\, +\, |5\rangle & \cr
}
;\lambda_- = {\rm e}^{\displaystyle -i\frac{2\pi}{3}}
\nonumber
\\\label{eigenv}\\
&\cases{
 |v_3\rangle\equiv\lambda_+|2\rangle\, +\, \lambda_-|3\rangle\, +\, |6\rangle & \cr
& \cr
 |v_4\rangle\equiv\lambda_-|1\rangle\, +\, \lambda_+|4\rangle\, +\, |5\rangle & \cr
}
;\lambda_+ = {\rm e}^{\displaystyle +i\frac{2\pi}{3}}
\nonumber
\end{eqnarray}

The other cyclic permutation, $P_{(3, 2, 1)}$, has the same eigenvectors, 
with the eigenvalues exchanged. 

By looking at the behavior of such states under the exchange 
of two particles, it is easy to find out that the two invariant subspaces
${\cal H}'_1$ and ${\cal H}'_2$ are generated respectively by 
$\{v_1, v_4\}$ and $\{v_2, v_3\}$. 
Therefore, the effect of a generic exchange is a rotation of the
vectors in ${\cal H}'_j$, with a change of the eigenvalues of $P_{(1, 2, 3)}$
and $P_{(3, 2, 1)}$. 
For example, the action of $P_{(2, 3)}$ on $v_1$ gives
\begin{equation} 
\pmatrix{0&1&0&0&0&0\cr1&0&0&0&0&0\cr0&0&0&1&0&0\cr
0&0&1&0&0&0\cr0&0&0&0&0&1\cr0&0&0&0&1&0}
\, \pmatrix{0\cr {\rm e}^{-i\frac{2}{3}\pi}\cr {\rm e}^{+i\frac{2}{3}\pi}\cr 0\cr 0\cr 1}\, 
=\, \pmatrix{{\rm e}^{-i\frac{2}{3}\pi}\cr 0\cr 0\cr {\rm e}^{
+i\frac{2}{3}\pi}\cr 1\cr 0}\, . 
\end{equation}
The resulting vector is $v_4$, which corresponds to a different 
eigenvalue of both cyclic permutations. The
same result applies for the other two exchanges. 

In the next Section we investigate the consequences of these properties 
of the cyclic permutations for the specific case of three nuclei bound 
in a rigid molecule, and the possible physical implication of the existence 
of these states not obeying SP. 

\section{Symmetry-forbidden states in symmetrical 3-nuclei
 molecules}
\label{sect:forbid}

\subsection{Three spin-0 identical nuclei}

The prototypal molecule that we want to consider is composed by three
identical spin-zero nuclei disposed at the vertices of an equilater
triangle, as sketched in Fig. \ref{d3}, whose symmetries belong to the dihedral
group $D_3$. Anyway, the discussion below can be extended to any
symmetrical molecule with additional nuclei in the plane
(point group $D_{3h}$). \par 

Within the Born-Oppenheimer approximation, the total wave function can
be decomposed in the usual way as
\begin{equation}
\Psi\, =\, \Psi_e \Psi_n \Psi_v \Psi_r\, , 
\end{equation}
where the partial wave functions are respectively the
electronic, nuclear spin, vibrational and rotational components. To
simplify our description, we can assume the molecule to be in the
ground electronic and vibrational state, with species $^1\Sigma$
(which is the most common case), or, in other words, we assume the
total electronic spin, electronic angular momentum and vibrational
angular momentum to be zero. Therefore, the electronic and vibrational
wave functions $\Psi_e$ and $\Psi_v$ are completely symmetric under any
permutations of the labels of the nuclei. In the case of spin-zero
nuclei also the nuclear spin wave-function $\Psi_n$ is symmetric under
permutations, and therefore we must consider the symmetry of the
rotational wave-function alone. \par

For this purpose, we briefly recall the main properties of the
rotational states of such a molecule, which are eigenstates of the
rigid rotational Hamiltonian
\begin{equation}
H\, =\, B({\bf
 J}/\hbar)^2\, -\, (B-C)({\bf K}/\hbar)^2\, , 
\end{equation}
where $B$ and $C$ are the two rotational constants, {\bf J} is the total
angular momentum and {\bf K} is its projection on the three-fold
symmetry axis perpendicular to the molecule's plane. The eigenvalues
of the Hamiltonian are of the
form
\begin{equation}
E(J, K)\, =\, BJ(J+1)\, -\, (B-C)K^2\, , 
\end{equation}
with $K=-J, -J+1, \cdots, J$, and are strictly degenerate in the sign
of $K$ as far as the molecule is symmetric, even if non-rigidity and
perturbations are considered (the degeneracy accounts for the
undistiguishability of the two orientations of {\bf K} with respect to
the molecular axis). 
The eigenfunctions are somewhat complicated
functions of $J$, $K$, and of the Eulerian angles \cite{herzberg}; 
they also depend parametrically on the equilibrium positions of
the three nuclei in the molecular potential. Here we
are interested in their transformation properties under a generic
rotation of the molecule. The general rule for such transformations
is the usual
\begin{equation}
\Psi_r\, \rightarrow\, \Psi_r
 e^{\displaystyle {i\alpha \bf{u_{\alpha}\cdot J}}}\, 
 , \label{rot1}
\end{equation}
where $\alpha$ is the angle of rotation and ${\bf u_{\alpha}}$ is the
versor of the plane containing $\alpha$. To find the symmetry
character of the rotational states, we want to compare this
transformation rule to those associated to the permutations of
particles. For such purpose we write the generic rotational states in
a formal way as a function of the coordinates of the three minima in
the molecular potential
\begin{equation}
|{\bf x_1, x_2, x_3}\rangle\, . 
\end{equation}
These states are defined in the 6-dimensional Hilbert space defined in
the previous Section. We note that a permutation of the quantum labels
corresponds to a classical permutation of the nuclei's mean positions, and
therefore all the permutations are equivalent to rotations of the
molecule. \par

We start by considering rotations in the plane
of the molecule by angles $\theta$=$\epsilon2\pi/3$, with
$\epsilon$=$\pm$1, which are equivalent to the cyclic permutations
$P_{(1, 2, 3)}$, $P_{(3, 2, 1)}$. Since the effect on the molecular
wave function of these rotations must be the same of
such cyclic permutations, we can compare the general transformation
rules under rotations and under permutations.
For the formers, the appropriate phase shift is given by
\begin{equation}
\Psi_r\, \rightarrow\, \Psi_r \exp^{\displaystyle
 {i\epsilon 2\pi K/3}}\, , 
\label{rot2}
\end{equation}
since the rotation is in the molecule's plane, while for the
permutations we have
\begin{eqnarray}
\Psi_r\, \rightarrow\, \Psi_r \qquad
{\rm and} \qquad 
 \Psi_r\, \rightarrow\, \Psi_r \exp^{\displaystyle
 {i\epsilon 2\pi/3}}\, , 
\label{rot3}
\end{eqnarray}
respectively for states in ${\cal H}_+$, ${\cal H}_-$ and in ${\cal
 H}'$. Therefore, a state with $K$=3$q$, with $q$ integer, can be defined in
${\cal H}_+$ or ${\cal H}_-$, while states with $K$=3$q\pm$1 can be
defined only in ${\cal H}'$. \par 

For the special case of a rotational state with $K$=0 we can consider
also rotations about any of the three symmetry axes in the plane of
the molecule by a angles $\phi$=$\epsilon\pi$. Such rotations are
equivalent to the three exchanges $P_{(1, 2)}$, $P_{(2, 3)}$ and
$P_{(1, 3)}$, and also the correspondent phase shifts in the wave
function must be identical. This class of rotations transform the wave
function as
\begin{equation}
\Psi_r\, \rightarrow\, \Psi_r \exp^{\displaystyle
 {i\epsilon \pi J}}\, , 
\label{rot4}
\end{equation}
according to Eq. (\ref{rot1}), since the angular momentum {\bf J} 
lays in the molecule's plane. The exchanges are instead
characterized by
\begin{eqnarray}
\Psi_r\, \rightarrow\, \Psi_r\qquad {\rm and} \qquad 
\Psi_r\, \rightarrow\, \Psi_r \exp^{\displaystyle
 {i\pi }}\, , 
\label{rot5}
\end{eqnarray} 
for states belonging respectively to ${\cal H}_+$ and ${\cal H}_-$. As
a result, the even-J states belong to ${\cal H}_+$, and the odd-J ones
to ${\cal H}_-$, as summarized in
Tab. \ref{subspaces}. 


From the results obtained above, it appears that all the rotational
states with $K$=3$q\pm$1 are strictly forbidden by the SP, which
requires any physical state to be defined in ${\cal H}_+$ or ${\cal
 H}_-$. Moreover, the odd-$J$ states with $K$=0 are forbidden by the
spin-statistics, since the identical nuclei are spin-0 particles. The latter
case is very similar to that of symmetrical molecules with two
identical nuclei \cite{deangelis,hilborn96,modugno}. These
consequences of both the SP and the spin statistics are well known in
the field of molecular spectroscopy \cite{herzberg}, since they lead
to the absence of more than two thirds of the rotational states in
molecules with the proper symmetry, as confirmed by the experimental
investigation performed so far. \par

We now discuss a property of the SP-forbidden states, which can
be defined in the unsymmetrical subspace ${\cal H}'$. Specifically, 
the states with $K$=3$q$-1 could be described by the vectors ${v_1, 
 v_2}$, while the ones with $K$=3$q$+1 by the vectors ${v_3, v_4}$ to
preserve the equality of the phase shifts reported in Eq. (\ref{rot2})
and Eq. (\ref{rot3}). As we noted, such pairs of vectors do not define
invariant subspaces, and therefore the physical states have to be
defined in the two invariant subspaces ${\cal H}'_1$ and ${\cal
 H}'_2$. Although in each of these subspaces the sign of the
rotational phase shift is not defined (see Eq. (\ref{eigenv})), also the sign
of $K$ is not a physical observable, due to the degeneracy of the
states with $\pm K$. In other words, even if in ${\cal H}'$ it is not
possible to define a complete set of operators due to its
multi-dimensionality, and in particular the sign of $K$ is undefined, 
no paradox arises, since we have no means to measure such a sign. \par

\subsection{Three spin-1/2 identical nuclei}


In case of a molecule containing three identical nuclei with non-zero
spin, we should consider also the influence of the nuclear spin
wave function; in the following we will discuss the special case of
spin 1/2. Since the spin is non-zero, we can now define a non-trivial
Hilbert space for the spin, in the same way we defined it for the
spatial coordinates. The total Hilbert space will be
the direct product of the rotational and spin spaces
\begin{equation}
{\cal H}\, =\, {\cal H}_S \otimes {\cal H}_R\, . 
\end{equation}
\par

The general properties of the Hilbert space can be easily found, 
considered the discrete nature of the quantum labels associated to each
particle. The generic three-particles state is now defined as
\begin{equation}
|S_{z1}, S_{z2}, S_{z3}\rangle\, , 
\end{equation}
and there are two states with total spin $I$=1/2 and one with $I$=3/2
(two doublet and one quartet). Using the formalism of Section II it
is possible to assign the quartet state to the symmetrical subspace
${\cal H}_+$, and the two doublet to the unsymmetrical ${\cal H}'$. 
We note that is not possible to build non-vanishing states completely
antisymmetric under permutations of more than two spin-1/2 particles, 
since each quantum label can assume only two values. \par

Even in absence of an electronic or vibrational angular momentum, the
nuclear spin can couple to the rotational angular momentum and give
rise to a very small hyperfine splitting of the rotational states. The
molecular states will therefore be eigenstates of the total angular
momentum {\bf F=J+I}. We want now to assign each molecular state to a
particular subspace of the overall Hilbert space, as we did in the
previous case. For such purpose we have to decompose the representation
of the S$_3$ group on the direct product Hilbert space 
${\cal H}_S \otimes {\cal H}_R$, into irreducible representations. 
By using general rules \cite{wu50} we easily obtain
the decomposition shown in 
Tab. \ref{moltiplication}. 


It is then straightforward to evaluate the appropriate Hilbert
subspace for the overall rotational and nuclear spin states, using the
associations of Tab. \ref{subspaces}; the results are reported in
Tab. \ref{spin}. If the hyperfine splitting of the rotational states
is not resolved, as it is usually the case, the states with $K\ne$ 0 are no more
completely forbidden by the SP, while the even-$J$ states with $K$=0
are forbidden by both the SP and the spin-statistics. Specifically, 
the states defined in ${\cal H}_+$ will have $I$=3/2, and the ones in
${\cal H}'$ $I$=1/2. As a consequence of the multidimensionality of
${\cal H}'$, it is impossible to distinguish between the two doublet states, 
but this is not a problem, since they are rigorously degenerate. 
In fact the invariant subspaces of ${\cal H}'$ are eigenstates of
$I_z$, and therefore also in presence of a magnetic field there is no
fundamental principle other than the SP to forbid the existence of
these unsymmetrical states. \par


We conclude by noting that in the case of a nuclear spin larger than
1/2 it is possible to build also antisymmetrical spin states in ${\cal
 H}_-$, and therefore none of the rotational states with $K$=0 is forbidden.

\subsection{Non-planar symmetrical molecules}

We now extend the discussion to non-planar symmetrical molecules
containing three spin-0 or spin-1/2 identical nuclei
(point group $C_{3v}$: the molecule has additional non-identical
nuclei out of the plane containing the identical nuclei). The main
difference from the planar case stands in the transformation rule of
the $K$=0 rotational states under rotations about a symmetry axis in
the plane of the three identical nuclei. These rotations, by angles
$\phi$=$\epsilon\pi$, are not identical to permutations of two
identical nuclei, as it was for a planar molecule, since also the
additional nuclei out of the plane has rotated. On the other
hand, the combination of such a rotation with an inversion of the
coordinates of all the nuclei with respect to the origin cannot be
distinguished from a permutation of two nuclei. \par

The properties of the rotational states under space inversion are well
known in the field of molecular spectroscopy \cite{herzberg}, and here
we give only the main result. If the non-rigidity of the molecule is
taken into account, each rotational state is split into two states
with definite symmetry under inversion ({\it s}- and {\it a}-species, 
respectively symmetric and antisymmetric under inversion). The
splitting of the degeneracy of these pairs of states is indicated as
inversion splitting. The analogous of the transformation rule of
Eq. (\ref{rot4}) for the rotation plus inversion is therefore
\begin{equation}
\Psi_r\, \rightarrow\, \pm\Psi_r \exp^{\displaystyle
 {i\epsilon \pi J}}\, , 
\label{rot6}
\end{equation}
where the positive (negative) sign refer to the {\it s}-species ({\it
 a}-species). If we compare this rule with the one for the exchange
of two particles (Eq. (\ref{rot5})) we can assign each rotational state
to the proper symmetry class, as summarized in Tab. \ref{nonplanar} for
the cases of spin-0 and spin-1/2. \par


In the last Section we will indicate one molecular species in which the
inversion splitting is sufficiently large to resolve the states
forbidden by the SP from those allowed.

\section{Testing small symmetry violations}
\label{sect:test}

As we noted above, those rotational states forbidden by the SP can be
defined consistently in the unsymmetrical subspace ${\cal H}'$, since
they are not violating any basic principle of Quantum Mechanics. The
discussion can be extended to particular classes of rotational
sub-states in excited vibrational and electronic states; due to the
different symmetry of the vibrational and electronic wave functions, 
different classes of $K$-levels and $J$-levels can be forbidden by the
SP. In any case, electromagnetic transitions between rotational levels
are allowed only if the involved states belong to the same symmetry
type, in accordance with the superselection rule. \par

Therefore, a violation of the SP can be thought of only in terms of
the appearance of a population of nuclei identical between themselves,
whose quantum states belong to ${\cal H}'$, and whose properties
(mass, spin, charge) are the same as those of the normal nuclei. The
presence of a non-zero population in SP-forbidden rotational states of
the molecules formed by such nuclei could be detected by exciting
electromagnetic transitions towards other forbidden states, in the
same way it was done in previous experiments for testing the
statistics. For a matter of sensitivity, the experimental
investigation should focus on states {\em completely} forbidden by the
SP alone, or by both the SP and the spin-statistics, since we expect
the violation, if present, to be very small. Therefore, the most
interesting states to be probed in both $D_{3h}$ and $C_{3v}$
molecules are those with $K$=3$q\pm$1 in the case of $S$=0, and those
with $K$=0 in the case of $S$=1/2. \par

We note also that all the statements above would not lose validity in
case of breakdown of the Born-Oppenheimer approximation. Indeed, the
symmetry character of the particles under permutation can only be
assigned to a single class ({\em i.e.} Hilbert subspace), as stated by
the super-selection rule, and no coherent superposition is allowed
\cite{caldirola}. Therefore, if the total wave function appears to be
correctly described as symmetrical (unsymmetrical) in the frame of the
Born-Oppenheimer approximation, it would be so for any degree of
violation of the approximation itself. The same argument applies in
case of presence of external perturbations, such as electric or
magnetic fields: even if they can change the molecular wave function
to a large extent in a continuous way, the symmetry character of the
wave function will be locked to a single class. \par

\section{Proposal for a
 high-sensitivity test of the symmetry on oxygen nuclei and protons}
\label{sect:proposal}

A simple tool to investigate for possible violations of the SP would
be high sensitivity spectroscopy of a thermal sample of gas, in search
for a non-zero absorption of light by symmetry-violating molecules, as
performed in previous experiments on the spin-statistics. It is
possible to define a few general criteria to choose the proper
molecule for such an experimental search. The first important
parameter is the weight of the molecule, which determines the spacing
between adjacent rotational transitions (the rotational constants $B$
and $C$ are inversely proportional to the mass), and therefore also
the capability of resolving the forbidden transitions from the allowed
ones \cite{modugno99}. This point is very important, since in
non-linear molecules the number of transitions which can possibly
interfere in the detection of SP-violations is increased with respect
to linear molecules, due to the increased molteplicity of rotational
states. Secondly, in order to have a high sensitivity in detecting a
violation of the SP, one should probe strong transitions starting from
low-laying rotational states in the ground electronic and vibrational
state, which have the largest occupation probability. As usual, in
this kind of molecules the strong electronic transitions are confined
to the UV, where laser sources are not easily available, and therefore
one should rely on the fundamental vibrational transitions in the
infrared, which are somewhat weaker. It is therefore important to find
a near coincidence between a vibrational band and the emission of a
coherent laser source suitable for high sensitivity spectroscopy. The
quantum-cascade semiconductor lasers are particularly interesting from
this point of view, since they can be designed to emit almost
everywhere in the mid-infrared, they are widely tunable and have
low-noise characteristics \cite{sharpe}.\par

In the case of spin-0 nuclei, the lightest molecule with the proper
symmetry is the plane SO$_3$ (group $D_{3h}$), whose vibrational
energies \cite{shimanouchy} are reported in Tab. \ref{vibr}. In
principle the sensitivity for detection of SP-violating transitions
can be expected to be comparable to those obtained in previous
experiments on molecular systems. It could possibly be reduced by
increased experimental difficulties (for example, SO$_3$ is not
chemically stable in presence of oxygen, and therefore may not be easy
to have the proper pressure and cell volume to optimize the
sensitivity) or more fundamental li\-mi\-ta\-tions, such as the
presence of weak hot-transitions close to the SP-forbidden ones.


An alternative approach could be to perform the spectroscopic
investigation on a supersonic beam of SO$_3$, instead of a thermal
sample. With this technique, the problems connected with the high
reactivity of SO$_3$ could possibly be solved, while the reduction in
density of the sample could be compensated by the reduction of the
rotational temperature and by an increase of the detection
sensitivity. In addition, one would benefit from the accompanying
increase in resolution, for a better identification of the detected
spectrum of transitions. \par

As for the case of spin-1/2 nuclei, it is possible to find a
relatively light plane molecule, BH$_3$ (group $D_{3h}$), on which to
search violations of the SP and of the spin-statistics for $K$=0
rotational states. As we mentioned, the same symmetry properties can
be found also for the corresponding rotational states of non-planar
symmetrical molecules (group $C_{3v}$) with inversion splitting. One
of the most interesting species of this kind is NH$_3$, which is
stable and particularly light, and has very strong vibrational
absorption bands in the infrared spectral region covered by
semiconductor lasers. Moreover, its spectrum has been the subject of
extensive investigation, and therefore the assignment of observed
lines can be performed in a relatively easy way. In addition, it is
characterized by a very large inversion splitting of the rotational
spectrum, and therefore the forbidden lines can be easily
distinguished form the allowed ones. For such molecules, which are
composed by spin-1/2 nuclei, we noted that for the $K$=0 states
different hyperfine components are forbidden by the SP and by the
spin-statistics. Since the hyperfine slitting of the rotational levels
is usually negligible (at least for $\Sigma$ electronic states
\cite{herzberg}), we can expect to resolve the two classes of substates
only thanks to Zeeman effects in strong magnetic fields. \par

To conclude, we note that in the case of BH$_3$
or NH$_3$, one would be testing the SP and the spin-statistics for
{\it protons}. Such an experimental test would be particularly
interesting to compare the present bound for violation of the
statistics for composed nucleons \cite{modugno} to a corresponding
bound for fundamental nucleons, also in view of the recent theoretical
predictions \cite{greenberg99}.

\section{Conclusions}
\label{sect:conlusion}

We have shown how to define consistently particular classes of
rotational levels of symmetrical molecules containing three identical
nuclei, by allowing for violations of the symmetrization postulate.
This violation, if existing, must be small, according to the results
of the past experimental observations. However, it is possible to look
with high sensitivity for the presence of tiny population in
SP-forbidden states, using spectroscopic tools in a scheme similar to
those of previous experiments on O$_2$ and CO$_2$.  If we allow for
the presence of molecules composed by SP-violating nuclei in a sample
of gas, which are identical to the normal molecules in all but the
symmetry, then such a measurement can be interpreted as a test of the
SP. This kind of experiment would represent a substantial improvement
with respect to the past investigations on possible violations of the
spin-statistics connection. We propose a few simple molecules
containing spin-0 nuclei (SO$_3$) or spin-1/2 nuclei (BH$_3$ and
NH$_3$) as candidates for a high sensitivity spectroscopic
investigation to be performed on infrared vibrational transitions,
with the help of semiconductor lasers. The latter species are
particularly interesting, since they represent systems on which to
test the SP for fundamental nucleons.\par

\acknowledgments

We acknowledge stimulating discussions with G. M. Tino.


\begin{table}[htbp] 
\begin{center} 
 
\vskip 12pt 
\begin{tabular}{c c c c} 
$K$ & $J$ &Subspace & Forbidden by\\ 
\hline 
$3q$, $q\ne$0 & any &${\cal H}_+$, ${\cal H}_-$ & - \\ 
$3q\pm$1& any &${\cal H}'$ & SP \\ 
0 & even &${\cal H}_+$ & - \\ 
0 & odd &${\cal H}_-$ & SS\\ 
\end{tabular} 
\caption{Appropriate Hilbert subspaces for rotational states of a 
 $D_{3h}$ molecule containing spin-0 identical nuclei, in the ground 
 electronic and vibrational state. The states completely forbidden by 
 the SP or by the spin-statistics (SS) are indicated. } 
\label{subspaces} 
\end{center} 
\end{table} 
 \begin{table}[htbp] 
\begin{center} 
 
\vskip 12pt 
\begin{tabular}{c | c c c c} 
&${\cal H}_+$ & ${\cal H}_-$ & ${\cal H}'$& \\ 
\hline\\ 
${\cal H}_+$ & ${\cal H}_+$ & ${\cal H}_-$ & ${\cal H}'$\\[4mm] 
${\cal H}_-$ & ${\cal H}_-$ & ${\cal H}_+$ & ${\cal H}'$\\[4mm] 
${\cal H}'$ & ${\cal H}'$ & ${\cal H}'$ & ${\cal H}_+\oplus{\cal H}_-
\oplus{\cal H}'$\\[4mm] 
\end{tabular} 
\caption{Decomposition of the direct product Hilbert space 
${\cal H}_S \otimes {\cal H}_R$, into symmetry inequivalent invariant
subspaces (under S$_3$).} 
\label{moltiplication} 
\end{center} 
\end{table} 

\begin{table}[htbp] 
\begin{center} 
 
\vskip 12pt 
\begin{tabular}{c c c c c} 
$K$ & $J$ & $I$ &Subspace & Forbidden by\\ 
\hline \\
3$q$, $q\ne$0 & any &1/2& ${\cal H}'$ & SP \\ 
3$q$, $q\ne$0 & any &3/2& ${\cal H}_+$, ${\cal H}_-$ & - \\[4mm] 
\hline\\  
$3q\pm$1& any &1/2& ${\cal H}_+$, ${\cal H}_-$, ${\cal H}'$ & - \\ 
$3q\pm$1& any &3/2& ${\cal H}'$ & SP\\[4mm] 
\hline\\  
0 & even &1/2& ${\cal H}'$ & SP \\ 
0 & even &3/2& ${\cal H}_+$ & SS \\ 
0 & odd &1/2& ${\cal H}'$ & SP\\ 
0 & odd &3/2& ${\cal H}_-$ & -\\[4mm] 
\end{tabular} 
\caption{Appropriate Hilbert subspaces for the rotational and
  spin states of a 
 $D_{3h}$ molecule containing  by spin-1/2 identical nuclei. } 
\label{spin} 
\end{center} 
\end{table} 
 \begin{table}[htbp] 
\begin{center} 
 
\vskip 12pt 
\begin{tabular}{c | c c c c} 
Nuclear& Inversion & $J$ &Subspace & Forbidden by\\ 
spin $S$&symmetry&&\\ 
\hline\\ 
&s & even &${\cal H}_+$ & - \\ 
0&s & odd &${\cal H}_-$ & SS\\ 
&a & even &${\cal H}_-$ & SS\\ 
&a & odd &${\cal H}_+$ & -\\[4mm] 
\hline\\ 
&s & even &${\cal H}_+$, ${\cal H}'$ & SS, SP \\ 
1/2&s & odd &${\cal H}_-$, ${\cal H}'$ & -\\ 
&a & even &${\cal H}_-$, ${\cal H}'$ & - \\ 
&a & odd &${\cal H}_+$, ${\cal H}'$ & SS, SP\\[4mm] 
\end{tabular} 
\caption{Appropriate Hilbert subspaces for $K$=0 rotational states of a 
 $C_{3v}$ molecule. In the case of $S$=1/2 each rotational state is
 composed by two states with different total nuclear spin $I$, and the
 finer assignment of the Hilbert subspaces can be done following Tab. \ref{spin}.} 
\label{nonplanar} 
\end{center}
\end{table}
 \begin{table}[h]
\begin{tabular}{c|ccccc}
Molecule & Spin &$\nu_1$ & $\nu_2$ & $\nu_3$ & $\nu_4$ \\[2mm]
\hline\\
SO$_3$ & 0 &1065 & 498 & 1391 & 530\\
BH$_3$ & 1/2 && 1125 & 2828 & 1640\\
NH$_3$ & 1/2 &3337 & 950 & 3444 & 1627 \\[2mm]
\end{tabular}
\caption{Vibrational energies (in cm$^{-1}$) and spin of the identical
 nuclei of selected molecules which present rotational states
 forbidden by the SP.}
\label{vibr}
\end{table}

\end{document}